\documentclass[a4paper,12pt]{article}
\usepackage{amsmath}
\usepackage{graphicx}              % Include figure files

\newcommand{\be}{\begin{equation}}
\newcommand{\ee}{\end{equation}}
\title{Gauss-Hermite Approximation Formula}
\author{Krzysztof Pomorski\\
 Katedra Fizyki Teoretycznej, University MCS, Lublin, Poland}
\date{}
\begin{document}
\maketitle
\begin{abstract}
A multidimesional function $y(\vec r)$ defined by a sample of points $\{\vec
r_i,y_i\}$ is approximated
by a differentiable function $\widetilde y(\vec r)$. The problem is solved by
using the Gauss-Hermite folding method developed in the nuclear shell
correction method by Strutinsky.
\end{abstract}

%%%%%%%%%%%%%%%%%%%%%%%%%%%%%%%%%%%%%%%%%%%%%%%%%%%%%%%%%%%%%%%%%%%%%%%%%%%%%%

\section{Introduction}
\label{intro}

Our aim is to approximate a sample of points $\{\vec r_i,y_i\}$ which
represents a measured or hard to evaluate data by a differentiable function
y(x). We would like to solve this problem using the Gauss-Hermite folding
method which idea was originally proposed by V.M. Strutinsky \cite{St66} and
later-on generalized in Ref.~\cite{NT69}. A detailed description of this method
may be found already in text-books, e.g. in Ref.~\cite{RS80}. Having the width
of the folding function comparable with the average distance between $x_i$
points in the $i^{\rm th}$ direction one can obtain the folded function which
goes very close to the data points but increasing its width one can also wash
out the fine structure stored in the data. Usually the Strutinsky method was
used to realize the second scope. The parameter of the folding procedure will
be determined by requirement that the integral in the $i^{\rm th}$ direction of
the folded function should be the same as the integral evaluated with 
$\{x_i,y_i\}$ pairs using the trapezium rule. A corresponding Fortran program
for the approximation in the $n$-dimensional space is listed in Appendix.

%%%%%%%%%%%%%%%%%%%%%%%%%%%%%%%%%%%%%%%%%%%%%%%%%%%%%%%%%%%%%%%%%%%%%%%%%%%%%%

\section{General folding formulae in the one-dimesional case}
\label{sec:1}

We consider an ensemble of $N$ points $\{ x_{i} \}$ distributed uniformly in
the interval $[ a, b ]$\footnote{Truly speaking the assumption about the 
uniform distribution of the points is too strong. It is sufficient to assume 
that the points $x_i$ have to cover the whole interval $[ a, b ]$ and to be 
ordered i.e. $x_{i+1} >x_i)$.}. To each point $x_{i}$ corresponds a point 
$y_{i}$, and we assume there exists a function $y(x)$ such that:
\be
  y_{i} = y(x_{i}) \,\,.
\ee

Let $j_{n}(x,x')$ be a symmetric function of its arguments 
(i.e. $j_{n}(x,x')=j_{n}(x',x)$) having the following properties:
\be
  \int\limits_{-\infty}^{+\infty} j_{n}(x,x')\,dx = 1 
\label{norm}
\ee
and
\be
  P_{k}(x) = \int\limits_{-\infty}^{+\infty} P_{k}(x')\,
                    j_{n}(x,x')\, dx' \,\,,
\label{Strut}
\ee
where $k \leq n$ are the even natural numbers and $P_{k}$ is an arbitrary 
polynomial of the order $k$. In the following the function $j_n(x,x')$ will  be
called the folding function of the $n^{\rm th}$ order. An example of such a
folding function can be a combination of the Gauss function and the Hermite 
polynomials of argument proportional to $|x-x'|$ frequently used in the 
Strutinsky shell correction method \cite{St66,RS80}. More detailed description 
of the Strutinsky folding function will be given in the next section.

With each discrete point $(x_{i},y_{i})$, one can associate the function 
$\widetilde{y}_{i}(x)$ defined by:
\be
\widetilde{y}_{i}(x) = \int\limits_{-\infty}^{+\infty} y_{i}\, \delta(x' - x_{i})\,
                  j_{n}(x,x')\, d x' \,\,,
\ee
where $\delta(x)$ is the Dirac $\delta$-function, so that 
\be
  \widetilde{y}_{i}(x) =  y_{i}\, j_{n} (x,x_{i}) \,\,.
\ee
Using Eq.~(\ref{norm}) it is easy to verify that the integral of the function 
$\widetilde{y}_{i}(x)$ is
\be
\int\limits_{-\infty}^{+\infty}\widetilde{y}_{i}(x)\,dx = y_{i} \,\,.
\ee
Let us construct the function $\widetilde{y}(x)$ by summing
up all functions $\widetilde{y}_{i}(x)$ corresponding to each $x_i$ point
\be
  \widetilde{y}(x) = \sum_{i=1}^{N} \omega_{i}\, \widetilde{y}_{i}(x) \,\,.
\ee
The Lebesgue theorem says that the function $\widetilde{y}(x)$ is an
approximation of $y(x)$ if the weights $\omega_{i}$ are determined from the
assumption that the integrals of the unfolded and folded functions are (nearly)
equal:
\be
  \int\limits_{a}^{b} y(x)\,dx \approx \int\limits_{-\infty}^{+\infty}
              \widetilde{y}(x)\,dx =  \sum_{i=1}^{N} \omega_{i}\, y_{i} \,\,.
\label{int}
\ee
The Riemann formula for the integral of the function $y(x)$ between bounds $a$
and $b$ reads:
\be
\int\limits_{a}^{b} y(x)\,dx = \lim_{N\to\infty}\, 
                        \sum_{i=1}^{N}  y(x_{i})\,\Delta x_{i} \,\,,
\label{rim}
\ee
where $\Delta x_{i}$ is set to: 
\be
  \Delta x_{i} = {1\over 2}\,(x_{i+1} - x_{i-1}) 
\ee
with $x_0=a$ and $x_{N+1}=b$. Comparing Eqs. (\ref{int}) and (\ref{rim}) one
can see that a reasonable choice of the weight is 
\be
\omega_{i} = \Delta x_{i} \,\,.
\ee
If the number $N$ of sample points $(x_{i},y_{i})$ large enough than
the condition (\ref{int}) will be fullfiled.\\

So, finally the folded function $\widetilde y(x)$ is given by
\be
\widetilde y(x) = \sum_{i=1}^N  y_i\,\Delta x_i\, j_n(x,x_i) \,\,.
\label{appy}
\ee

%%%%%%%%%%%%%%%%%%%%%%%%%%%%%%%%%%%%%%%%%%%%%%%%%%%%%%%%%%%%%%%%%%%%%%%%%%%%%%

\section{Gauss-Hermite folding function}
\label{sec:2}

Let the folding function $j_n(x,x')$ be a modified Gauss function
\be
 j_n(x,x') = \frac{1} {\gamma\sqrt{\pi}} \,
             {\rm exp}\left\{-\left({x-x'\over\gamma}\right)^2\right\}
             f_n\left(\frac{x-x'} {\gamma}\right)\,, 
\label{jn}
\ee
where $\gamma$ is the parameter and $f_n({x-x'\over\gamma})$ is the so called
correction polynomial of the order $n$ determined by the Strutinsky condition 
(\ref{Strut}). In the following we would like to evaluate the coefficients of
the correction  polynomial using some properties of the Hermite polynomials
which are orthogonal with the weight equal to the Gauss function.

Let us introduce the variable $ u = (x - x')/\gamma$ which belongs to the
interval $(-\infty ,+\infty )$. The smearing function $j_n(x,x')$ and the
polynomial $P_n(x)$ (\ref{Strut}) can be now written as
\be 
 j_n(x,x') = \frac{{\rm e}^{-u^2}} {\gamma\sqrt{\pi}} \,f_n(u) \,,
\ee

\be
  P_n(x') =  P_n(x - \gamma\, u) \equiv   {P_n}'(u)\,, 
\ee
and
\be
  P_n(x) =  P_n(x + \gamma\, 0) \equiv   {P_n}'(0)\,.
\ee
Let us decompose the function ${P_n}'(u)$ into series of the Hermite polynomials
$H_{i}(u)$
\be
 {P_n}'(u) = \sum^{n }_{i=1}a_{i}\,H_{i}(u)\,\,.
\label{hep}
\ee
Now the condition (\ref{Strut}) for $k=n$ can be written as
\be
 {P_n}'(0) = \frac{1}{\sqrt{\pi}} \int\limits^{+\infty }_{-\infty } \,
 {P_n}'(u){\rm e}^{-u^{2}} \,f_n(u)\, du 
\ee
and inserting the relation (\ref{hep}) one obtains 
\be
 \sum^{n}_{i=1}a_{i} \left\{\frac{1}{\sqrt{\pi}} \int^{+\infty}_{-\infty}
  {\rm e}^{-u^{2}} \, H_{i}(u) \, f_n(u)\,du - H_{i}(0)\right\} = 0 \,\,.
\ee
The last equation should be fullfiled for arbitrary values of $a_i \neq 0$) 
what leads to the following set of equations
\be
 \frac{1}{\sqrt{\pi}} \int^{+\infty}_{-\infty} \, {\rm e}^{-u^2} \, H_{i}(u) \,
f_n(u)\,du
 = H_{i}(0) \,\,,
\label{coef}
\ee
where $i=0,2,...,n$. From the other side the correction function $f_n(u)$ can 
be also decomposed into series of the Hermite polynomials
\be
 f_n(u) = \sum^{n}_{k=1}C_{k}H_{k}(u)\,\,.
\label{corr}
\ee
Inserting the above relation to Eq.~(\ref{coef}) one obtains
\be
H_i(0) = \sum^{n}_{k=1}C_{k} \frac{1}{\sqrt{\pi}} \int^{+\infty}_{-\infty} \,
   {\rm e}^{-u^{2}}H_{i}(u) \, H_{k}(u)\,du \,\,.
\label{rown}
\ee
Then using the orthogonality properties of the Hermite polynomials
\be
\int^{+\infty}_{-\infty} \, {\rm e}^{-u^{2}}H_{i}(u) \, H_{k}(u)\,du =
       2^{i}\,i!\,\delta_{ik} \,\,,
\ee
one obtains the coefficients of the correction polynomial (\ref{corr})
\be
  C_{i} = {1\over 2^{i}i!}\,H_{i}(0)
\ee
The values of the Hermite polynomials at zero-point are
\be
  H_i(0) = \left\{2^n
 \begin{array}{cl}
         1            & \mbox{for}~i=0        \,\,,\\
  (-1)^n (2n-1)!!~~~~ & \mbox{for}~i = 2n     \,\,, \\
       0              & \mbox{for}~i = 2n + 1 \,\,,
\end{array}\right.
\label{zero}
\ee
so
\be
 C_i = \left\{
 \begin{array}{cl}
           1                         & \mbox{for}~i=0         \,\,,\\
(-1)^n \frac{(2n-1)!!}{2^n(2n)!}~~~~ & \mbox{for}~i = 2n > 0  \,\,,\\
           0                         & \mbox{for}~i = 2n + 1  \,\,.
\end{array}\right.
\label{Ci}
\ee

The first few coefficients $C_i$ and the corresponding Hermite polynomials 
are:\\
\be
\begin{array}{ll}
C_0 = 1			         & H_0 = 1  			        \\
&\\
C_2 = - {1\over 4}		 &  H_2(u) =  4u^2 - 2  	        \\
&\\
C_4 = + {1\over 32}		 &  H_4(u) = 16u^4 -  48u^2 + 12        \\
&\\
C_6 = - {1\over 384}~~~~~~~~~~~  &  H_6(u) = 64u^6 - 480u^4 + 
                                                           720u^2 - 120 \\ 
\end{array}
\ee
and the corresponding correction polynomials have the following form
\be
\begin{array}{l} 
  f_0(u) = 1                                                          \,\,, \\
 ~\\
  f_2(u) = \frac{3}{2}  -              u^2                            \,\,, \\
 ~\\
  f_4(u) = \frac{15}{8} -  \frac{5}{2} u^2 + \frac{1}{2} u^4          \,\,, \\ 
 ~\\
  f_6(u) = \frac{35}{16}- \frac{35}{8} u^2 + \frac{7}{4} u^4 - \frac{1}{6}u^6 
     \,\,, \\
\end{array}
\label{fcorr}
\ee

Finally the function $\widetilde y(x)$ approximated by the Gauss-Hermite folding 
reads:
\be
\widetilde y(x) = \frac {1}{\gamma\sqrt{\pi}}\,\sum_{i=1}^N  y_i\,\Delta x_i \,
    {\rm exp}\left\{-\left({x-x_i\over\gamma}\right)^2\right\}
    \,f_n\left({x-x_i\over\gamma}\right) \,\,.
\ee

As a rule the smearing parameter $\gamma$ is arbitrary and it can be
different at each point $x_i$. But it should be related to the distance $\Delta
x_i$ between subsequent points if one would like to approximate the function
stored in the mesh of $\{x_i,y_i\}$ points. Similarly one has to choose the 
$\gamma$ parameter of the order of the period-length of the fine structure 
(e.g. shell effects) in case when one would like to wash out this structure 
from the function $y(x)$.
\begin{figure}
\begin{center}
\resizebox{0.60\textwidth}{!}{%
  \includegraphics{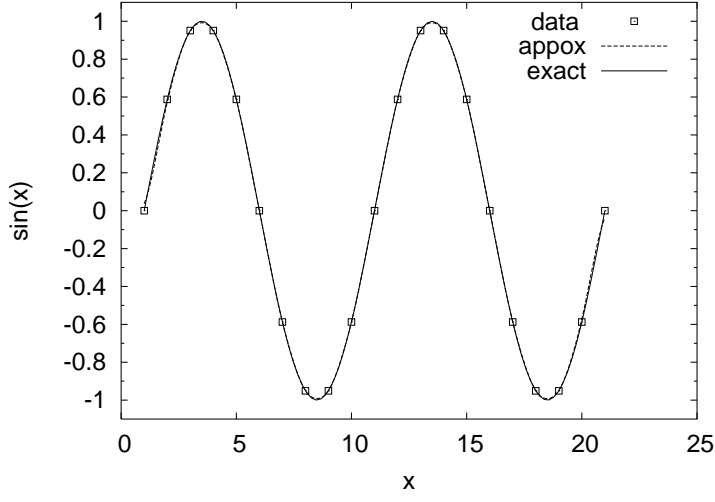}}
\end{center}
\caption{The sinus function and its approximation with the $2^{nd}$ order
Gauss-Hermite smoothing.}
\label{fig:1}       % Give a unique label
\end{figure}

%%%%%%%%%%%%%%%%%%%%%%%%%%%%%%%%%%%%%%%%%%%%%%%%%%%%%%%%%%%%%%%%%%%%%%%%%%%%%%%%

\section{Multidimensional case}
\label{sec:3}

The extension of the formalism described in the previous sections is
straightforward. Let assume that the data points are stored in a m-dimensional
array $Y[1:N_1,1:N_2,...,1:N_m]$ which corresponds to the ordinates given by 
the $m$ one-dimensional arrays $X_i[1:N_i]$, where $i=1,2,...,m$. It means 
that each element of $Y[i_1,i_2,....,i_m]$ corresponds to the coordinate
$X_k[i_k]$ with $k=1,2,...,m$.

The ensemble of the Hermite polynomials $H_i(x)$ forms a complete basis of 
orthogonal functions in which an arbitrary m-dimensional function 
$F(\vec r)\equiv F(x_1,x_2,...,x_m)$ can be expanded
\be 
  F(x_1,x_2,...,x_m) = \sum_{i_1=0}^\infty\,\sum_{i_2=0}^\infty ....
  \sum_{i_m=0}^\infty C_{i_1i_2...i_m}\,H_{i_1}(x_1)\,H_{i_2}(x_2)\,...\,
  H_{i_m}(x_m) \,\,.
\label{exp}
\ee

The same is true for any polynomial $P_n(x_1,x_2,...,x_m)$ of the order $n$ but
in this case the upper limit of the sums in equation analogous to (\ref{exp})
will be $n$. It means that the folding function in the m-dimensional space is 
simply the product of the $m$-th one-dimensional foldings performed in each 
single direction:
\be
J_n(x_1,x_1';x_2,x_2';....;x_m,x_m')= \prod_{i=1}^m j_n(x_i,x_i') \,\,.
\ee
The folded function $\widetilde Y(x_1,x_2,...,x_m)$ is given by the
equation analogous to (\ref{appy}):
\be
\begin{array}{rl}
\widetilde Y(x_1,x_2,\,...\,,x_m) =& \sum\limits_{i_1=0}^\infty\Delta x_{i_1}
       \sum\limits_{i_2=0}^\infty\Delta x_{i_2} \,...\,
       \sum\limits_{i_m=0}^\infty\Delta x_{i_m}\,\cdot\, \\
        & \\
   \cdot& Y[i_1,i_2,\,...\,,i_m]\,j_n(x_1,X_1[i_1])\,...\, j_n(x_m,X_m[i_m])\,\,.
\end{array}
\label{appY}
\ee
The folding function $j_n(x_i,x_i')$ for the Gauss-Hermite smoothing (\ref{jn})
is 
\be
 j_n(x_i,x_i') = \frac{1} {\gamma_i\sqrt{\pi}} \,
             {\rm exp}\left\{-\left({x_i-x_i'\over\gamma_i}\right)^2\right\}
             f_n\left(\frac{x_i-x_i'} {\gamma_i}\right)\,\,. 
\label{jn1}
\ee
The Gaussian width $\gamma_i$ can be different in each coordinate. The
correction polynomial $f_n$ was already given by Eq.~(\ref{corr}) and it
is tabulated in (\ref{fcorr}) for n=0,2,4, and 6.

%%%%%%%%%%%%%%%%%%%%%%%%%%%%%%%%%%%%%%%%%%%%%%%%%%%%%%%%%%%%%%%%%%%%%%%%%%%%%%

\section{Some data illustrating the quality of the method}
\label{sec:4}
   
The second order ($n=2$) Gauss-Hermite folding in a four-dimensional space is
used. Taking into account higher order correction polynomials (\ref{fcorr}) one
does not increase significantly the quality of the approximation of the
function in the middle of the data region but it would need a more careful
treatment of the  problem at the edges.  The folding is performed using the $p$
mesh points closest to the given point in each direction. The tested function
are spanned on $21^4$ points. In addition it is assumed that outside the the
data region $\{x_i,y_i\}$ the function which should be folded has a constant
value (equal to the value of the first or the last point in the given direction
depending from which side of the data region one takes the data for folding).

The data in Table~\ref{Tab1} are listed for  some
values of the smearing parameter $\gamma$ in order to see its influence on the
accuracy of the approximation. The cosines function in the four-dimensional
space is chosen as the test function:
\be
Y(x_1,x_2,x_3,x_4)=cos(r) \,\,, 
\ee
where $r=\sqrt{x_1^2+x_2^2+x_3^2+x_4^2}$ and the equidistant points 
$x_n(i) \in \langle -2\pi,2\pi\rangle$ for 
$i=1,2,...,21$ and $n=1,2,3,4$. The upper and lower limit are $Y_{max}=1$ and
$Y_{min}=-1$ respectively.
\begin{table}[h]
\caption{The approximation of the four-dimensional $\cos(r)$ function by the 
$2^{nd}$ order Gauss-Hermite folding on basis of $p=5$ and $p=7$ closest to 
the given points in function of the smearing parameter $\gamma$.}
\label{Tab1}
\begin{center}
\begin{tabular}{|c||c|c|c||c|c|c|}
\hline
$1/\gamma$&$\delta_{avr}$&$\delta_{\rm min}$&$\delta_{\rm max}$
          &$\delta_{avr}$&$\delta_{\rm min}$&$\delta_{\rm max}$\\
\cline{2-7}
&\multicolumn{3}{|c||}{$p=5$}&\multicolumn{3}{|c|}{$p=7$}\\
\hline
0.98  & 0.0081  &  -0.0296  &   0.0530  &  0.0030  &  -0.0073  &  0.0241  \\
1.00  & 0.0072  &  -0.0261  &   0.0485  &{\bf 0.0029}&{\bf -0.0074}&{\bf 0.0242}\\
1.02  & 0.0065  &  -0.0233  &   0.0452  &  0.0030  &  -0.0076  &  0.0249  \\
1.04  & 0.0060  &  -0.0210  &   0.0428  &  0.0032  &  -0.0080  &  0.0260  \\
1.06  & 0.0057  &  -0.0192  &   0.0414  &  0.0035  &  -0.0086  &  0.0276  \\
1.08  &{\bf 0.0057}&{\bf -0.0179}&{\bf 0.0409}&  0.0040  &  -0.0093  &  0.0295  \\
1.10  & 0.0059  &  -0.0171  &   0.0411  &  0.0046  &  -0.0102  &  0.0320  \\
\hline
\end{tabular}
\end{center}
\end{table}

The root mean square deviation 
\be
\delta_{\rm avr} = \left(\frac{\sum\limits_{i=1}^N \{Y[\vec r(i)]-
       \widetilde Y[\vec r(i)]\}^2}{N-1}\right)^{1/2} 
\ee
as well as the maximal in plus difference ($\delta_{\rm max}$) and the minimal
in minus one  ($\delta_{\rm min}$) are evaluated for the $N=149057$ mesh and
inter-mesh points (in the middle) excluding the points which lie on two outer
layers (i.e two first or last rows or columns). Such a choice of the test nodes
was made in order to eliminate the influence of the border condition on the
deviation $\delta_{\rm avr}$. 

Some other examples of the accuracy of the $2^{nd}$ order Gauss-Hermite
approximations are listed in Table~\ref{Tab2}. The function written in the
first column are tabulated at 21 equidistant points in the each coordinate in
the 4-dimensional space in the interval written in the $2^{nd}$ column. The
smearing parameter $\gamma=0.93$ or $\gamma=1$ is chosen in case of the $p=5$
or $p=7$ point basis used when folding, respectively.
\begin{table}[h]
\caption[TT]{A few examples of the approximation accuracy in the
             four-dimensional space.}
\label{Tab2}
\begin{center}
\begin{tabular}{|c|c|c|c|c|c|c|c|}
\hline
Function $Y$&Range&$Y_{\rm min}$&$Y_{\rm max}$&$p$&$\delta_{\rm avr}$&
$\delta_{\rm min}$&$\delta_{\rm max}$\\
\hline

$\sin(r)/r$&$-2\pi:2\pi$&-0.2172&1&5& 0.0011 &  -0.0029 &  0.0128 \\
\cline{5-8}
          &            &       &   &7& 0.0005 &  -0.0012 &  0.0059 \\
\hline
$x_1^2+x_2^2+x_3^2+x_4^2$&-2 : 2& 0    & 8&5& 0.0053 &  -0.0218 &  0.0102 \\
\cline{5-8}
          &            &       &   &7& 0.0013 &  -0.0017 &  0.0014 \\
\hline
$(x_1\cdot x_2\cdot x_3\cdot x_4)^2$&-2 : 2& 0&256&5& 0.0076 &-0.2491 &0.1161\\
\cline{5-8}
          &            &       &   &5& 0.0017 &  -0.0191 &  0.0102 \\
\hline
$x_1\cdot x_2\cdot x_3\cdot x_4$&-2 : 2&-16&16&5&0.0014&-0.0180&0.0180 \\
\cline{5-8}
          &            &       &   &7& 0.0001 &  -0.0017 &  0.0017 \\
\hline 
\end{tabular}
\end{center}
\end{table}
It is seen in Tables~\ref{Tab1} and \ref{Tab2} that in all considered 
cases the root mean square deviation ($\delta_{avr}$) is of the order $10^{-3}$ 
or less of the maximal difference $Y_{max}-Y_{min}$ between the data points.

%%%%%%%%%%%%%%%%%%%%%%%%%%%%%%%%%%%%%%%%%%%%%%%%%%%%%%%%%%%%%%%%%%%%%%%%%%%%%%%%

\section{Summary and conclusions}
\label{sec:5}

A new method of the smooth approximation of a function defined on a sample
of points in a multidimensional space is proposed. The folding of the discrete 
data points using the Gauss-Hermite method of Strutinsky is performed.
Depending on the width of the Gauss function the folded function can be very
close to the approximated data or can give its average behavior only.

The folded function and all its derivatives are continuous. This significantly
increases the range of applicability of the method. Our approximation of the
discrete data can be used e.g. when solving transport equations or other type
of equations of motion in a multidimensional space, what is a frequent problem
in economy, meteorology and environment protection problems as well as in
molecular or nuclear dynamics.

The proposed approximation of the data can be also used in the computer
graphic art. It can wash out the fine structure from a photography keeping 
unchanged its average background. One can also think about the use of the new 
folding method when one evaluates the cross-sections of a multidimensional 
data which one has e.g. in the X-ray tomography. 

%%%%%%%%%%%%%%%%%%%%%%%%%%%%%%%%%%%%%%%%%%%%%%%%%%%%%%%%%%%%%%%%%%%%%%%%%%%%%%%

%%%%%%%%%%%%%%%%%%%%%%%%%%%%%%%%%%%%%%%%%%%%%%%%%%%%%%%%%%%%%%%%%%%%%%%%%%%%%%%%
\newpage

\section{Appendix}
The source of the fortran program for the $2^{nd}$ order Gauss-Hermite
approximation:  
\begin{center}
  \includegraphics[height=22cm, angle=0]{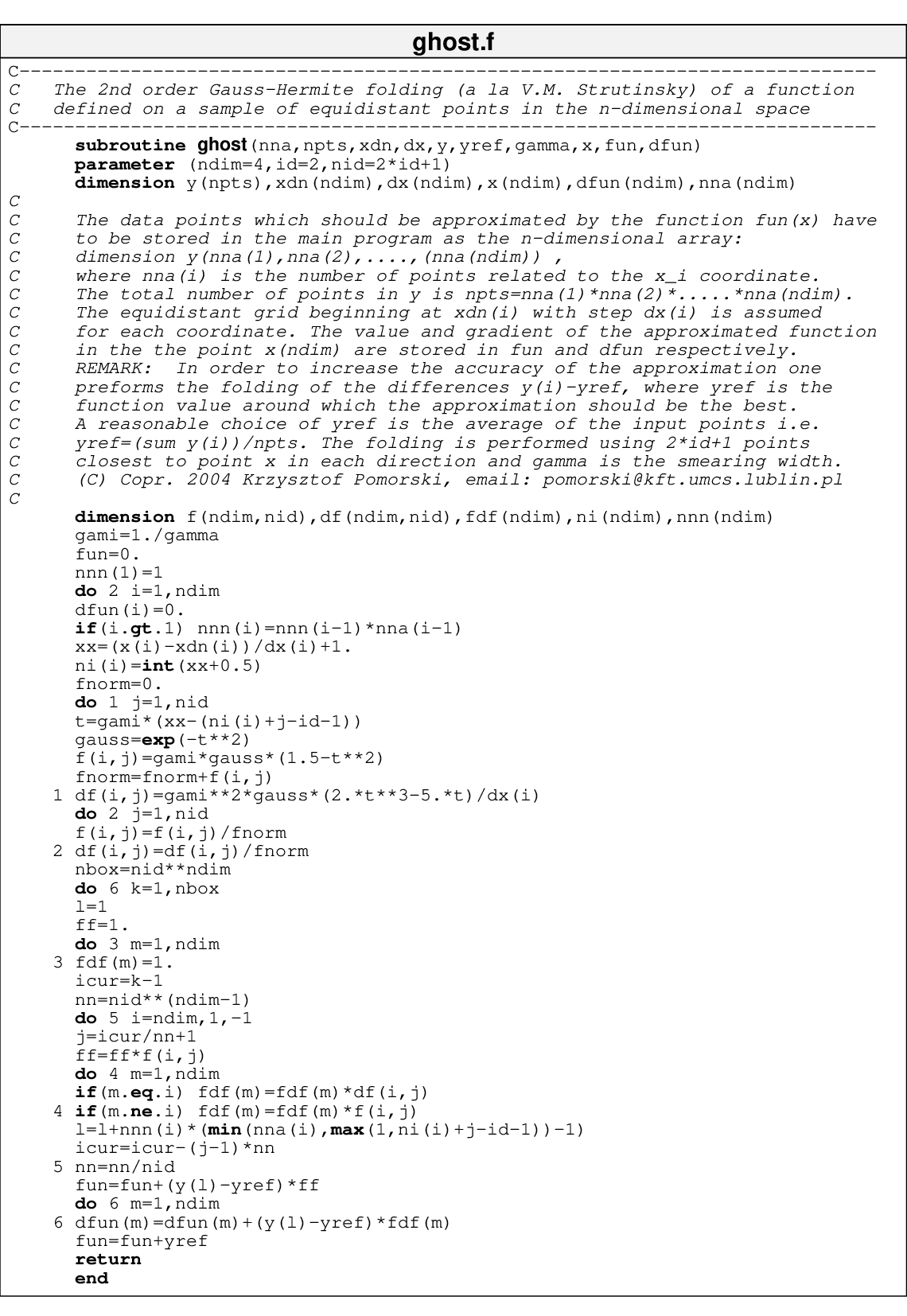}
\end{center}

\end{document}